\documentclass{article}
\usepackage[T1]{fontenc} 
\usepackage[utf8]{inputenc} 
\usepackage{ismir,amsmath,cite,url}
\usepackage{graphicx}
\usepackage{color}
\usepackage{multirow}
\usepackage{xurl}
\usepackage{bbold}

\usepackage{lineno}

\title{From Real to Cloned Singer Identification}

\multauthor
{Dorian Desblancs\hspace{0.5cm}Gabriel Meseguer-Brocal\hspace{0.5cm}Romain Hennequin\hspace{0.5cm}Manuel Moussallam}
{Deezer Research, Paris, France\\
{\tt\small research@deezer.com}
}

\def\authorname{D. Desblancs, G. Meseguer-Brocal, R. Hennequin, and M. Moussallam}

\usepackage[bookmarks=false,pdfauthor={\authorname},pdfsubject={\papersubject},hidelinks]{hyperref}

\begin{document}

\maketitle

\begin{abstract}

Cloned voices of popular singers sound increasingly realistic and have gained popularity over the past few years.
They however pose a threat to the industry due to personality rights concerns.
As such, methods to identify the original singer in synthetic voices are needed.
In this paper, we investigate how singer identification methods could be used for such a task.
We present three embedding models that are trained using a singer-level contrastive learning scheme, where positive pairs consist of segments with vocals from the same singers.
These segments can be mixtures for the first model, vocals for the second, and both for the third.
We demonstrate that all three models are highly capable of identifying real singers.
However, their performance deteriorates when classifying cloned versions of singers in our evaluation set.
This is especially true for models that use mixtures as an input.
These findings highlight the need to understand the biases that exist within singer identification systems, and how they can influence the identification of voice deepfakes in music.

\end{abstract}

\section{Introduction}\label{sec:introduction}
In April 2023, the track ``Heart on my Sleeve'' by an anonymous TikTok user Ghostwriter977 put the music industry in a frenzy \cite{nyt, josan2024ai}.
The artist used artificial intelligence (AI) based cloning technologies to turn their voice into Drake and the Weeknd's \cite{billboard}, two of the most popular singers in the world.
The song became very popular across music streaming platforms, before being removed by demand of the original artists' right owners.
This situation raised the need for singer identification systems that can also identify the original singer a synthetic voice was generated from.

In this paper, we train three embedding models for singer identification using a singer-level contrastive learning scheme, where positive pairs consist of segments with vocals of the same singers whilst negatives come from different singers.
These samples can be mixtures for the first model, vocals for the second, and both for the third.
The models are then evaluated on real singers using novel splits of two open datasets, the Free Music Archive (FMA) \cite{fma_dataset, fma_challenge} and MTG-Jamendo (MTG) \cite{bogdanov2019mtg}, and a closed dataset consisting of 176,141 songs that span 7500 popular singers.
We use this dataset due to its scale and the fact that its singers are often the target of music voice deepfakes, some of which we use in this paper.
We demonstrate that all three models are highly capable of classifying real voices, though genres that use effects on vocals, such as hip-hop, pop, and electronic music, and singers with long discographies can be much harder to classify.
We then test whether the performance of our models generalizes to cloned voices of singers present in our closed dataset, using songs from YouTube.
In this context, singers are often cloned onto famous instrumentals, or instrumentals that differ greatly from their usual environments.
We find that the performance of all three models deteriorates quite significantly.
This is especially true for models that use mixtures as inputs.
We hope that these findings can be useful for future singer identification works.
We believe that these should aim to design systems that can identify both a singer's real and synthetic voice, in the hopes of combating the growing problem of voice deepfakes in music.

We summarize the contributions of this work as follows: 1) We evaluate singer identification systems on songs with real singers and cloned voices of some of the same singers. 2) We offer a detailed inspection of their performance, and demonstrate that these systems struggle to classify synthetic voices, genres where audio effects are applied to natural voices, and singers with long discographies. For synthetic voices, this decline is even greater when instrumental information is present during training, and highlights the need to understand the biases that exist within singer identification systems. 3) We open source singer identification splits of two open datasets, the FMA and MTG, that can serve as future performance benchmarks for the task of singer identification in polyphonic mixtures.\footnote{\scriptsize\url{https://github.com/deezer/real-cloned-singer-id}}

\section{Related Work}

Singer identification, has been a staple of the music information retrieval (MIR) community for more than twenty years \cite{kim2002singer, fujihara2005singer}.
Early approaches aimed to attenuate the instrumental parts of a song through the use of vocal melody or pitch extraction and voice re-synthesis and detection algorithms \cite{mesaros2007singer, fujihara2010modeling, lagrange2012robust}.
Classic features, such as mel-frequency cepstral coefficients (MFCCs),
were then computed on these signals and used as inputs for a classifier.
The improvements in music source separation \cite{hennequin2020spleeter, defossez2019music} then led researchers to build models that classify singers using the vocals of each song \cite{sharma2019importance, hsieh2020addressing}.
More recently, self-supervised methods that process the vocal stem of each track have been shown to be effective for singer identification \cite{yakura2022self, torres2023singer}.
However, source separation is computationally costly.
As such, these algorithms are hard to deploy on catalogues that span millions of tracks.
Several works have attempted to build embedding models for singer identification that use mixtures as inputs \cite{lee2019learning, kim2021learning}, most notably by using triplet learning where anchors and positives come from the same singers in different instrumental environments.

The work in this paper focuses on testing whether singer identification systems trained on real voices generalize to cloned voices of the same singers.
This task must not be confused with the task of singing voice deepfake detection, which has very recently emerged in the signal processing community \cite{xie2024fsd, zang2024singfake}.
Both works introduce datasets for Chinese singing voice spoofing detection,
and demonstrate that state-of-the-art speech deepfake detectors fail to accurately predict whether the songs in their datasets are deepfakes.
After supervised training, their performance is improved.
However, \cite{zang2024singfake} also finds that the classifiers are not robust to unseen singers, languages, or musical contexts, suggesting the need for more complex methods.

Finally, audio embeddings learned using artist-based sampling scheme have been used in \cite{royo2018disambiguating}. The authors used metric learning, with anchors and positives coming from the same artists, to train a neural network for artist disambiguation.
More recently, \cite{alonso2022music} used sampling at the artist level for contrastive learning and downstream tasks such as genre and mood classification or music tagging.

\section{Experimental Setup}\label{sec:methods}

In this section, we first present the datasets used throughout this paper.
We then present the setup used to train an embedding model for singer disambiguation using contrastive learning.
Finally, we present how this model is used for singer identification.

\subsection{Datasets}\label{subsec:data}
We collect a vast number of popular, commercial, and annotated songs for both training and evaluating the embedding models.
The data and their singer annotations come from four sources: Deezer, MusicBrainz, Wikidata, and Discogs \cite{kong2024strada}.
The latter three are publicly available.
In total, we collect more than four million tracks that span $\sim2.6$ million artists.
We then filter out all tracks that are not comprised of vocal segments at least 75\% of the time.
For this, we use a simple deep learning model that classifies three-second segments into either an instrumental class or a vocal class across all songs.
We then filter out all unique singers that do not have at least two tracks.
This leaves us with 37,525 singers.
7500 of the ones that have at least seven tracks are used for our singer identification task.
The remaining 30,025 are used to train and validate our embedding models using contrastive learning.

\bgroup
\def\arraystretch{1.}%
\begin{table}[h!]
\centering
\begin{tabular}{ |c|c|c|c| } 
\hline
\textbf{Dataset} & \textbf{No. Singers} & \textbf{No. Songs} & \textbf{Songs/Singer} \\
\hline
Train & $25929$ & $181989$ & $\geq2$ \\ 
Validation & $4096$ & $8192$ & $2$ \\
\hline
Closed & $7500$ & $176141$ & $\geq7$ \\
FMA & $1019$ & $11676$ & $\geq5$ \\ 
MTG & $572$ & $7710$ & $\geq5$ \\
\hline
Cloned & $67$ & $377$ & N.A. \\
\hline
\end{tabular}
\caption{Attributes of each dataset used in this paper. The train and validation sets are used for training the embedding models. Upon initial collection, validation singers can have more than two mostly-vocal tracks; we however randomly select a segment with vocals from two tracks to keep the set constant. The closed, FMA, and MTG datasets are used for real singer identification. The cloned dataset contains songs collected from YouTube in which synthesized voices of real singers are used. The original singers in this dataset are present in the closed dataset.}
\label{tab:datasets}
\end{table}
\egroup

We then gather 377 tracks, from YouTube, with cloned voices from 67 singers in our closed dataset.
These are used to test our embedding models on music voice deepfakes.
We also test our models on two open music tagging datasets for real singer identification: the FMA and MTG.
For these, we first gather their artist tags.
We then filter out songs that are not comprised of segments with vocals at least 50\% of the time.
Artists with less than five songs are also removed.
This leaves us with 1019 artists for the FMA and 572 artists for the MTG.
Unlike the commercial, closed dataset, each song can contain more than one singer.
We however postulate that the trends observed in the results are highly indicative of our models' performances on the singer identification task.
We publish the subsets of data we used on these datasets for reproducibility and to serve as future benchmarks.
To the best of our knowledge, other open singer identification datasets, such as the VocalSet \cite{wilkins2018vocalset} and M4Singer \cite{zhang2022m4singer}, only contain snippets of a capella singing voices.
We hope future singer identification systems will also be evaluated on our proposed, more authentic musical data: singers singing to an instrumental.
Table \ref{tab:datasets} displays the attributes of each of these sets of data.

\subsection{Singer-Level Contrastive Learning}\label{subsec:contrastive}

We train the embedding models in a contrastive learning way to predict whether two songs are from the same singer.
During each training iteration, we begin by drawing a batch of $B = 128$ positive pairs, which correspond to pairs of segments with singing.
These pairs are drawn on the fly from different songs of the same singer.
For our Mixture model, these segments come from songs' mixtures.
For our Vocal model, these segments come from the vocal stem generated by Demucs \cite{demucs2022hybrid, defossez2021hybrid}.
Finally, for our Hybrid model, these segments are randomly sampled from either; the following positive pairings are possible during the contrastive learning task: vocal-vocal, vocal-mixture, mixture-vocal, and mixture-mixture.
This is done to better disambiguate singing voices, without the need for source separation during the downstream singer identification task.
All segments are sampled at 16000 Hz and have a six-second duration.
We then compute their mel-spectrograms and pass these through the small version of the transformer model from \cite{ast21b_interspeech}.
We use a FFT size of 800, a hop length of 400, and a total of 128 mel bins for our mel-spectrogram operation.
The transformer model then maps the resulting $128 \times 240$ tensor to an embedding of size 2048.
Similarly to \cite{chen2020simple, spijkervet2021contrastive, saeed2021contrastive}, these embeddings are passed through a fully-connected projector head.
In our case, this head maps our embeddings to outputs of dimension 2048, 1024, and 2048, and uses batch-normalization \cite{ioffe2015batch} and ReLU activations.
Let us denote the resulting projections by $y_i$, where $i \in [1, 2B]$.
For each positive pair $(i, j)$, we compute, and aim to minimize, the normalized temperature-scaled cross-entropy, or NT-Xent \cite{chen2020simple}, loss function, defined as:
\begin{equation}
    \ell_{i,j}
     = -\log \frac{\exp\left(\frac{1}{T}S(y_{i}, y_{j})\right)}{\sum_{k=1}^{2B} \mathbb{1}_{[k \neq i]} \exp\left(\frac{1}{T}S(y_{i}, y_{k})\right)},
    \label{eq:contra}
\end{equation}
where the indicator function $\mathbb{1}_{[k \neq i]}$ evaluates to 1 iff $k \neq i$. 
$T$ is a temperature parameter that helps the model learn from hard negatives.
In our case, we set $T = 0.2$ following \cite{wang2021understanding}.
$S(u, v)$ denotes the cosine similarity between vectors $u$ and $v$.
The average loss value is then backpropagated to the model.
We use the ADAM optimizer \cite{kingma2014adam} throughout training, with an initial learning rate of 0.0001.
This value is decreased by a factor of 0.5 at every 25-epoch validation loss plateau.
Note that one epoch corresponds to 32 training and validation iterations.
We stop training when the validation loss has plateaued for 100 epochs.

\subsection{Singer Identification}\label{subsec:singer_id}
We then train classifiers with the same architecture as the projector head from the previous section upon the frozen transformer models.
We evaluate these classifiers on two sets of data for real singer identification: the FMA and MTG open datasets and the closed set.
For all of these, we randomly set aside one track for testing and one track for validation per singer.
These are constant throughout all our experiments for reproducibility purposes.
Note that four segments from each validation track are selected randomly at the beginning of each singer identification experiment to keep the validation set constant.
At least three tracks per singer are then used for training.
During each training iteration, we select a segment with vocals on the fly to construct batches of size 100.
We minimize a Cross Entropy loss \cite{bishop2006pattern} using the ADAM optimizer with an initial learning rate of 0.01.
This value is decreased by a factor of 0.1 every 10-epoch validation loss plateau.
Here, one epoch is, again, equal to 32 training and validation iterations.
We stop training when the validation loss has plateaued for 20 epochs.
For each dataset's test tracks, the final singer prediction is obtained using a majority vote scheme, where each segment with vocals is passed through the frozen embedding model and classification head.
The singer with the most ``votes'' is then used as the track's final output.

We report all our results using 10 runs.
For both open datasets, we report results using all singers.
On the other hand, for our 7500-singer closed set, we report results from 100 to 1000 classes.
During each run, we randomly sample a subset of singers, on which we then train and evaluate a classifier.
Finally, for the cloned singers dataset, we: 1) train models to classify 100 to 1000 singers using the closed dataset; 67 of these are cloned singers, whilst the remainder are randomly sampled from the rest of our closed dataset. 2) try to classify the cloned singer of our deepfake tracks. Our goal is to evaluate whether singer identification systems trained on real singer data can correctly classify the singers' voice deepfakes.


\section{Results}\label{sec:results}

\subsection{Open Datasets}\label{subsec:open-source}
\bgroup
\def\arraystretch{1.}%
\begin{table}[h!]
\centering
\begin{tabular}{ |c|c|c|c| } 
\hline
\textbf{Dataset} & \textbf{Model} & \textbf{Top-1 Acc.} & \textbf{Top-5 Acc.} \\
\hline
\multirow{4}{4em}{\textbf{FMA}} & CLMR & $73.2$ +/-- $0.6$ & $73.6$ +/-- $0.6$ \\ 
 & Mixture & $76.6$ +/-- $0.5$ & $84.1$ +/-- $0.6$ \\ 
 & Hybrid  & $77.6$ +/-- $0.3$ & $85.1$ +/-- $0.6$ \\
 & Vocal   & $79.9$ +/-- $0.4$ & $85.7$ +/-- $0.3$ \\
\hline
\multirow{4}{4em}{\textbf{MTG}} & CLMR & $67.9$ +/-- $1.1$ & $68.0$ +/-- $1.1$ \\ 
 & Mixture & $78.5$ +/-- $0.5$ & $88.4$ +/-- $0.6$ \\ 
 & Hybrid  & $79.3$ +/-- $1.1$ & $88.7$ +/-- $0.6$ \\
 & Vocal   & $83.2$ +/-- $0.6$ & $91.3$ +/-- $0.6$ \\
\hline
\end{tabular}
\caption{Singer identification results obtained on the open datasets (\%). For each dataset, we report the top-1 and top-5 accuracies generated by the three models we train using singer-level contrastive learning. We also use the embeddings from \cite{meseguer2024}, called CLMR, as a baseline. These embeddings are trained in a similar fashion to \cite{spijkervet2021contrastive}, but on $\sim4$M tracks, and are used for both training and testing our classification heads to generate these results. We display the means and standard deviations over 10 runs.}
\label{tab:open_results}
\end{table}
\egroup

The results obtained on open datasets can be visualised in Table \ref{tab:open_results}.
One can immediately notice that the CLMR \cite{meseguer2024} results are inferior to the singer-level embedding models' results by at least a few percentage points.
On the FMA dataset, we notice a 3.4\% top-1 gap between the CLMR and Mixture models.
This gap grows to more than 10 percentage points using a top-5 accuracy and is even more exacerbated on the MTG dataset and with the Hybrid and Vocal models.
This highlights the fact that sampling at the singer-level is much more adapted than classic, high-performing self-supervised learning methods for pre-training a model for singer identification.
We observe these gaps even though the contrastive model from \cite{meseguer2024} is trained on more than 20$\times$ more tracks than the embedding models we trained for this paper.

\begin{figure*}[h]
    \centering
    \includegraphics[width=.91\textwidth]{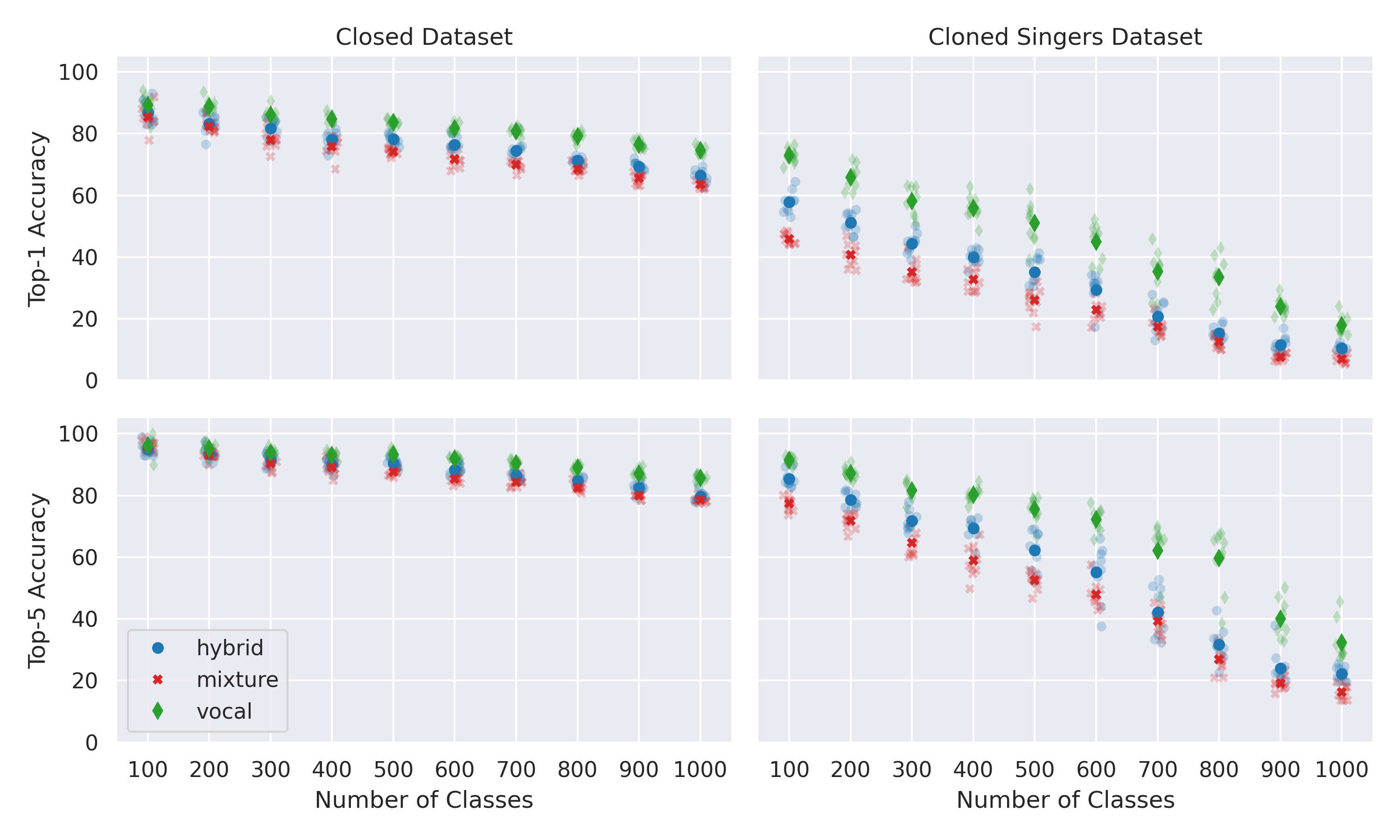}
    \caption{Singer identification results obtained on the closed (left) and cloned (right) datasets.
    We display results over 10 runs for 100 to 1000 singer classes.
    For each number of classes, we display the top-1 and top-5 accuracies for each run (pale markers), and the mean results between all runs (prominent markers).
    On the closed dataset, we randomly sample a subset of the 7500 singers on every run and display results on their test tracks.
    For the cloned dataset, we train our models to classify the 67 cloned singers and other randomly selected singers.
    We then display the results on the 377 spoofed tracks.}
    \label{fig:all_results}
\end{figure*}

We can also notice that the Vocal model outperforms the models that use mixtures as an input by a few percentage points.
More specifically: for the MTG dataset, we observe a 3.9\% gap compared to the Hybrid model on top-1 accuracy and a 2.6\% gap on top-5 accuracy.
The gap is less pronounced on the FMA data.
We can also notice that the Hybrid model, which samples both mixtures and vocal stems during pre-training, is slightly better-performing than the mixture model, though the performance gap never exceeds a percentage point.
This highlights one of our main findings: separating vocals from the rest of the track clearly helps our models disambiguate singers between each other.
However, we can obtain good performance using mixtures too.
In the realm of production, where source separation can be costly memory and time-wise, the results obtained using the Hybrid or Mixture models may suffice; they may not justify the need to separate vocal stems beforehand.
The results obtained on the closed dataset in the next section further emphasize this idea.

\subsection{Closed Dataset}\label{subsec:closed}
The results obtained on the closed dataset can be found on the left side of Figure \ref{fig:all_results}.
We observe a similar trend to the one observed on the open datasets: the Vocal model outperforms both models that use mixtures as inputs by a few percentage points.
Then, the Hybrid model outperforms the Mixture model, though the gap is narrow.
For example, 
for 400 classes, we observe mean top-5 accuracies of $89.2$\% 
for the Mixture model, $90.1$\% 
for the Hybrid model, and $93.2$\% 
for the Vocal model.
For 700 classes, we observe mean top-1 accuracies of $69.8$\% 
for the Mixture model, $74.5$\% 
for the Hybrid model, and $80.9$\% 
for the Vocal model.
As the number of classes grows, the gaps in performance are more pronounced, especially on the top-1 accuracy metric.
We however suggest that the gap between the Vocal model and models that work on mixtures does not warrant the need for source separation in production-like environments for real-singer identification.

\bgroup
\def\arraystretch{1.}%
\begin{table}[h!]
\centering
\begin{tabular}{ |c|c|c|c|c| } 
\hline
\textbf{No. Singers} & \textbf{Method} & \textbf{Dataset} & \textbf{Top-1} & \textbf{Top-5} \\
\hline
\multirow{2}{4em}{\textbf{300}} & \cite{kim2021learning} & MSD & $39.5$ & $69.2$ \\ 
 & Mixture & Closed & 78.0 & 90.4 \\ 
\hline
\multirow{3}{4em}{\textbf{500}} & \cite{lee2019learning, yakura2022self} & MSD & $47.9$ & $71.2$ \\
 & \cite{yakura2022self} & MSD & $63.1$ & $82.2$ \\ 
 & Mixture & Closed & $74.2$ & $87.6$ \\
\hline
\end{tabular}
\caption{Singer identification results for the same number of singers in this and previous works of the field (\%).}
\label{tab:comparison}
\end{table}
\egroup
Comparing our results to previous works in the field of music singer identification is quite difficult.
These report results on private datasets \cite{yakura2022self} or on the Million Song Dataset (MSD) \cite{Bertin-Mahieux2011, lee2019learning, kim2021learning}, a dataset whose audio is not publicly available.
That is why we hope future works in singer identification will also be evaluated on the open splits we report results on in Section \ref{subsec:open-source}.
We however report our worst and previous works' results for the same number of singers in Table \ref{tab:comparison}.
These highlight that our methodology is at the very least on par with previous works
and validate sampling at the singer level for contrastive learning when the downstream task is singer identification.

\begin{figure}[t]
    \centering
    \includegraphics[width=0.46\textwidth]{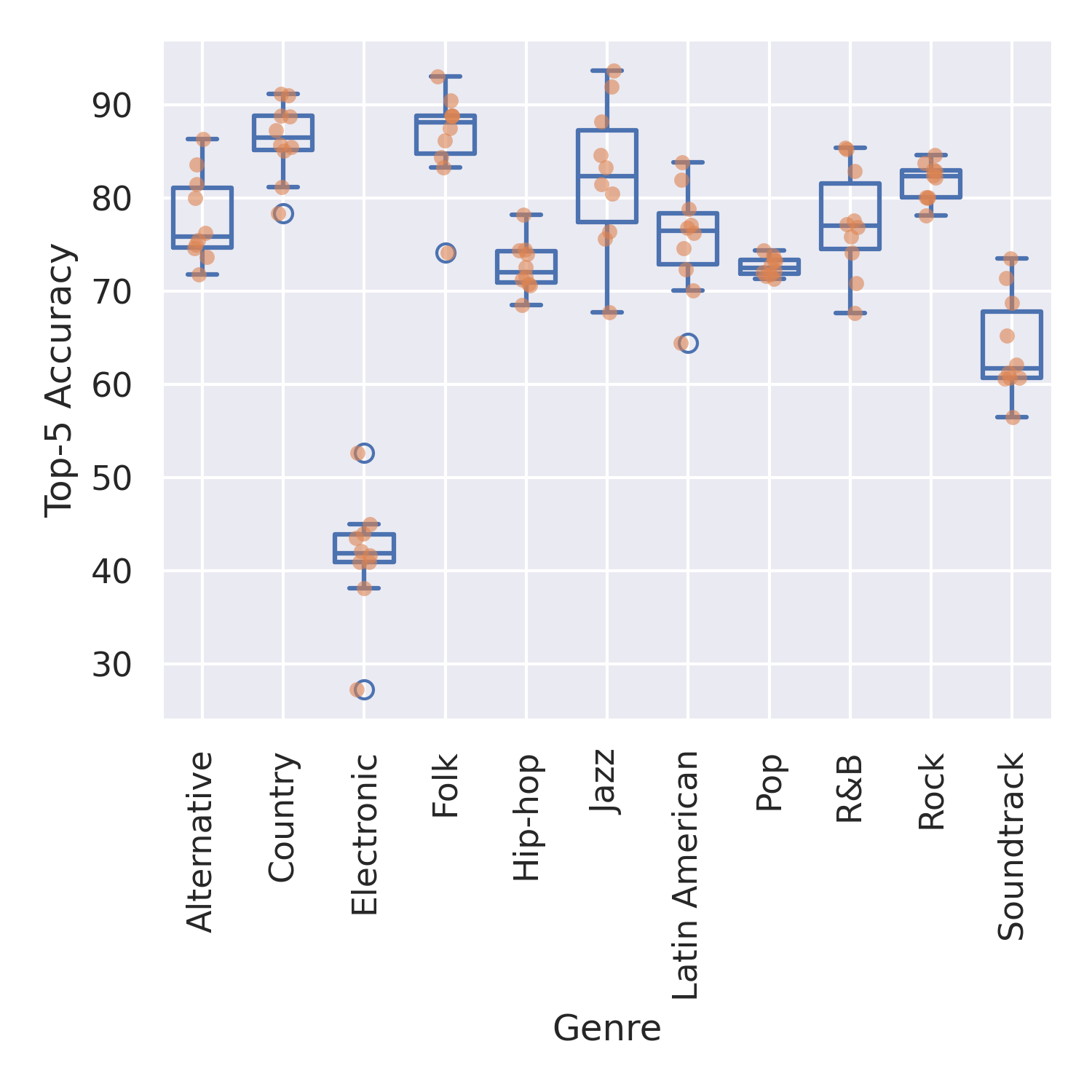}
    \caption{Vocal model performance by genre when trying to classify 1500 singers. The macro genre tags are gathered from Deezer and are unique for each test track. We display the mean top-5 accuracy for each run with the orange dots. The boxes then display the median and interquartile range (IQR) between runs. The whiskers extend to points that lie within 1.5 IQRs of the lower and upper quantiles. Finally, outlier runs have circles drawn around them. Genres containing less than 100 test tracks are omitted from this plot.}
    \label{fig:genres}
\end{figure}

We should however point out that, even though our models identify real singers quite well, there remain open challenges.
As displayed in Figure \ref{fig:genres}, our performance over musical genres is not uniform.
For example, for Country and Folk music, the mean top-5 accuracies are $86.3$\% 
and $86.6$\%.
On the other hand, for Hip-hop and Pop music, the mean top-5 accuracies are $72.7$\% 
and $72.7$\%.
The performance drops even further for Electronic music, where we observe a mean top-5 accuracy of $41.6$\%.
The same trends can be observed for top-1 accuracy, our other models, and different numbers of classes.
Hip-hop, Pop, and Electronic genres tend to employ effects such as reverb and vocoder on singing voices.
These effects can change a voice's timbre quite substantially, and seem to have an effect on our singer identification performance.
On the other hand, Folk and Country tend to have natural-sounding singing voices.
We suggest that future singer classification works should aim to lessen the gap between these genres, perhaps by introducing augmentations during either the embedding or classifier's training.
We also did experiments on the influence of language on performance, and did not find our results to be biased towards any of these.
We found all 10, commonly-represented languages to have a median top-5 accuracy between 68 and 82\% for 1500-singer identification, with substantial overlaps in distribution.

\begin{figure}[t]
    \centering
    \includegraphics[width=0.34\textwidth]{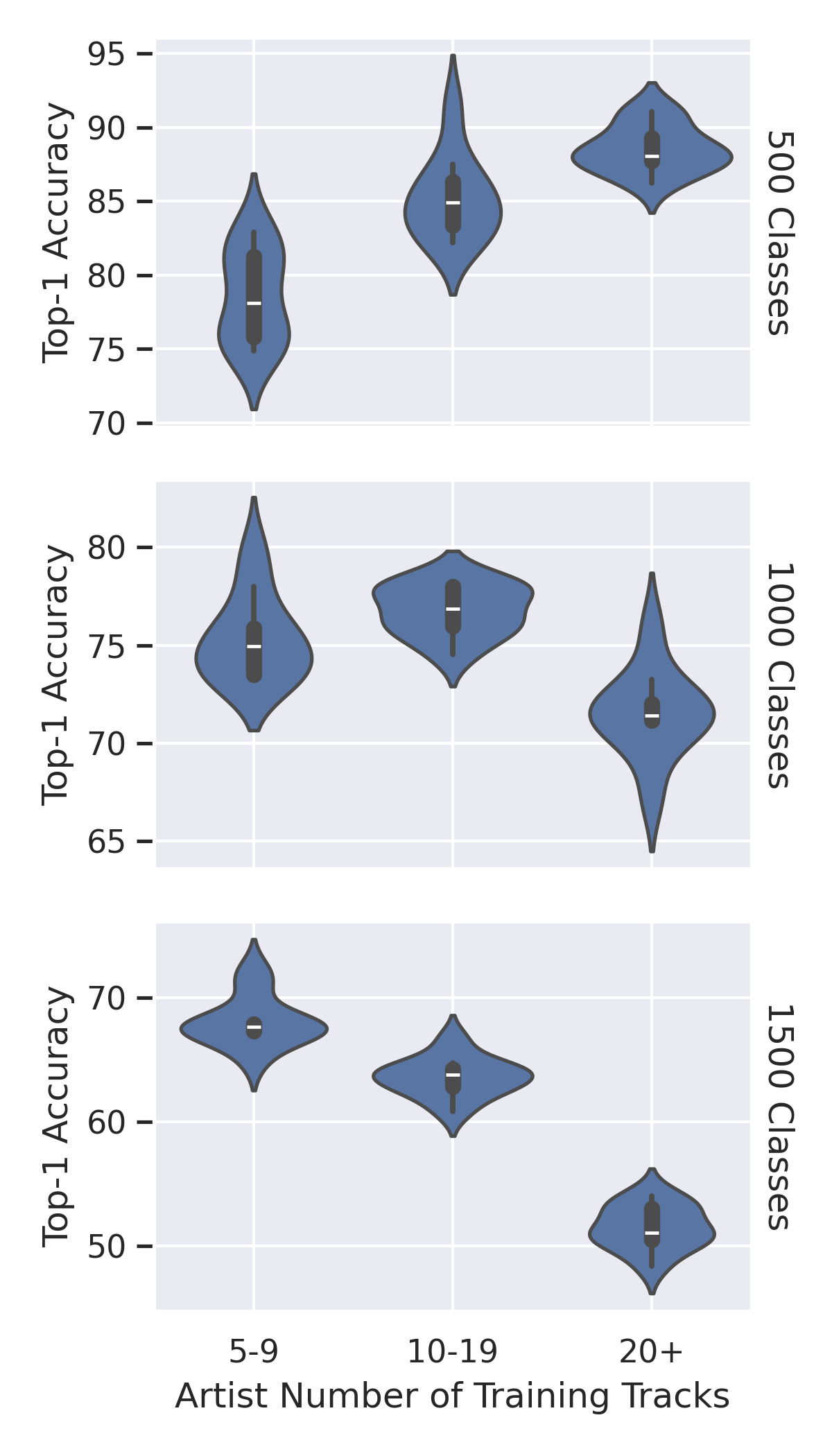}
    \caption{Vocal model performance over 500, 1000, and 1500-singer identification. We report results from each run in buckets that describe the number of training tracks per singer, that are used to train our classifiers. In the first, we display the top-1 accuracies observed for singers with only 5 to 9 training tracks. In the second, we display the top-1 accuracies observed for singers with 10 to 19 training tracks. Finally, in the last, we display the top-1 accuracies observed for singers with 20 or more training tracks. We report results using violin plots, where, for each bucket, the inner figure is a box plot similar to that in Figure \ref{fig:genres} and the outer figure is a kernel density estimation of the data.}
    \label{fig:num_train_tracks}
\end{figure}

One can also notice the following trend from Figure \ref{fig:num_train_tracks}: when we are trying to classify a small number of singers, having more tracks per singer for training leads to higher performance; on the other hand, when we are trying to classify a large number of singers, having fewer tracks for training leads to higher performance.
For example, for 500-singer identification, we merely observe a top-1 accuracy of $78.5$\% 
when singers have 5 to 9 training tracks.
This top-1 accuracy grows to $88.5$\% 
when singers have 20 or more training tracks.
On the other hand, for 1500-singer identification, we observe top-1 accuracies of $68.1$\% 
and $51.5$\% 
for these same buckets.
These trends can also be observed on our other models and for top-5 accuracy.
They suggest that, as the singer identification task gets harder, singers with more songs to their name, and most likely much longer careers, get harder to correctly classify than singers with just an extended play (EP) or album to their name.
This could be due to changes in style, mixing effects, or even singing voice.
We hope that future works in the field will design systems that are more robust to singing voice evolution over a variety of musical projects.



\subsection{Cloned Voices}\label{subsec:ai}
The results on our cloned dataset can be found on the right side of Figure \ref{fig:all_results}.
One can immediately notice a sharp decline between the performance we observed on real singers and synthetic ones.
For 200-singer identification, the worst-performing model on real singers, the Mixture one, has a mean top-1 accuracy of $82.3$\%.
In comparison, the best-performing model on synthetic voices, the Vocal one, has a top-1 accuracy of $65.8$\%.
For 600-singer identification, their respective top-5 accuracies are $85.3$\% 
and $72.3$\%.
The decline in performance on cloned singers is hence quite dramatic.
However, in a lot of ways, it is to be expected.
Synthetic voices can sometimes be quite unrealistic depending on the voice conversion or generation techniques used, which should obviously lead to deterioration in singer identification performance.

The more striking decline is that which we observe between models themselves.
More notably, the Vocal model performs substantially better than the models that use mixtures as inputs.
Starting at 100 classes, the mean top-1 accuracy of the Vocal model is $65.8$\% versus $51.2$\% for the Hybrid model and $40.1$\% for the Mixture model.
For 800 classes, we observe accuracies of $33.4$\% versus $15.3$\% and $12.7$\%.
On the one hand, for real singers, we found the gap in performance between the Vocal and Hybrid to be minimal enough to justify using mixtures over vocal stems, and hence avoid using source separation pre-processing.
Here, however, the answer is much more clear cut: the Vocal model is the only one with decent performance on cloned singer identification task, whilst the models that use mixture inputs see a very significant drop in performance.

\begin{figure}[t]
    \centering
    \includegraphics[width=0.47\textwidth]{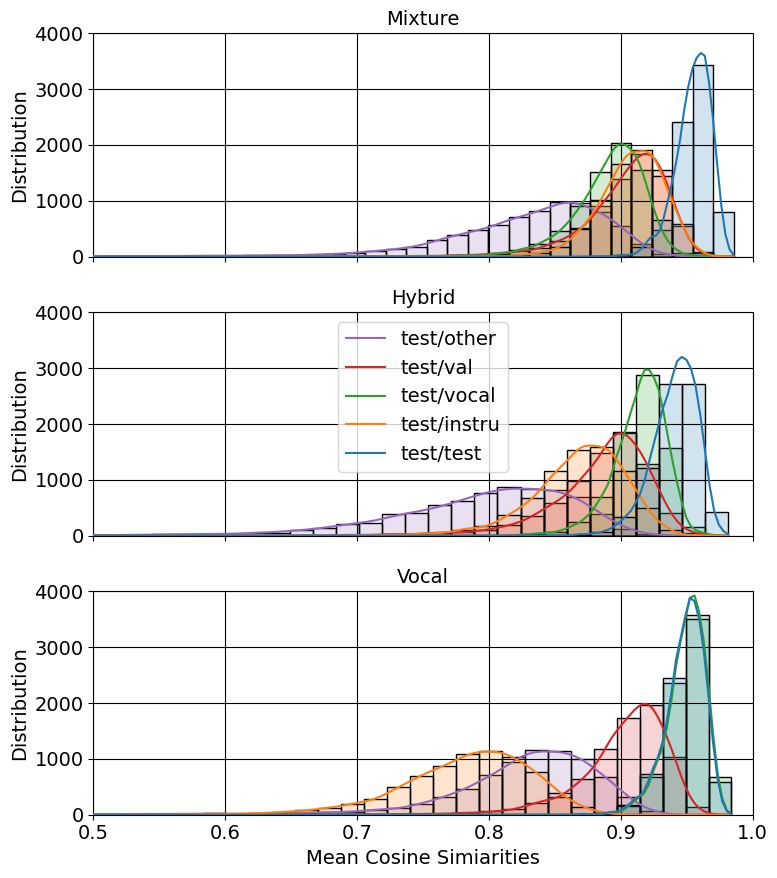}
    \caption{Mean all-pairs cosine similarity between each of the closed set singers' test track embeddings and: in purple (test/other), the embeddings from a random track from another singer; in red (test/val), their validation track embeddings; in green (test/vocal), their test track's vocal stem embeddings; in orange (test/instru), their test track's instrumental stem embeddings; in blue (test/test), the other embeddings from the same track. All embeddings are generated on segments with vocals.}
    \label{fig:cosine}
\end{figure}

The reason behind the performance drop between models is illustrated in Figure \ref{fig:cosine}.
When comparing the embeddings of each closed set test track to other embeddings of the same track, we see that these are very similar, with cosine similarities of $\sim95$\%.
However, the comparison with the test tracks' stem embeddings can differ significantly.
For the Mixture model, we see that the instrumental embeddings are actually more similar to the ``ground truth'' test track embeddings (GTEs) than the vocal embeddings, with mean similarities of $90.8$\% and $89.1$\%.
Even worse, the instrumental embeddings are closer to the GTEs than the validation track's.
Hence, even though our Mixture model, like the Hybrid and Vocal models, is pre-trained to disambiguate singers, we find that its embeddings are more suitable for finding similar songs based on instrumental information than vocal information.
This problem is partly solved in the Hybrid model and fully solved in the Vocal model.
Note that these results extend to other vector similarity measures such as Euclidean distance.

These findings outline why the models that are trained using mixtures drop off significantly on spoofed versions of famous artists.
On these tracks, singers are often used on an instrumental which is either from another famous track, or an instrumental which is very different from their usual environment.
Some of the cloned tracks' instrumentals are even present in their original tracks during training on real singers, which leads to obvious misclassifications.
As such, models that bias singers towards certain types of musical backgrounds fail to correctly identify them in altered contexts.
Source separation allows us to better disambiguate voices only during training, and thus classify synthetic versions of performers.
In the future, perhaps reintegrating mashups to alter a singer's context on the fly, such as was done in \cite{lee2019learning}, could lead to more robust singer identification models. These could solve the two main remaining problems in the field: 1) the need for source separation pre-processing and 2) the identification of cloned versions of existing singers.




\section{Conclusion}

In this paper, we train three models using singer-level contrastive learning.
The first is only trained using mixtures, the second is only trained using vocal stems, while the third is trained using both.
We find that all three models are highly capable of classifying real singers, though there remain open challenges, such as classifying genres that use more vocal effects and singers with long discographies.
However, all three models' performance decreases drastically when trying to identify cloned voices of existing singers.
This decrease is much more pronounced for models that are trained using mixtures.
These models bias singers towards certain types of instrumentals.
They therefore struggle to correctly classify them in different background music environments, such as those offered by singing voice deepfakes.
By publishing our results and novel, singer identification splits of the FMA and MTG datasets, we aim to generate more research in this field of MIR.
Future works could notably incorporate cloned voices in a few-shot fashion in the hopes of minimizing the gap between real and synthetic singer identification.


\section{Ethics Statement}

Our work offers a glimpse into how we, as a field, can identify the original singers in music voice deepfakes.
It is important that outputs of systems like ours not be used as justification to make important decisions, however, such as content removal from platforms.
As demonstrated in this paper, singer identification systems are often wrong; they often return false positives.
This is even more true on deepfakes.
As such, human emotion and decision-making should still be at the heart of the music deepfake battle.
Creative, talented singers should never see their work de-platformed because a machine learning model falsely said so.
The outputs of these models should always be interpreted with caution, as an indication but not a truth.


\bibliography{ISMIR2024_template}

\begin{thebibliography}{10}
\providecommand{\url}[1]{#1}
\csname url@samestyle\endcsname
\providecommand{\newblock}{\relax}
\providecommand{\bibinfo}[2]{#2}
\providecommand{\BIBentrySTDinterwordspacing}{\spaceskip=0pt\relax}
\providecommand{\BIBentryALTinterwordstretchfactor}{4}
\providecommand{\BIBentryALTinterwordspacing}{\spaceskip=\fontdimen2\font plus
\BIBentryALTinterwordstretchfactor\fontdimen3\font minus \fontdimen4\font\relax}
\providecommand{\BIBforeignlanguage}[2]{{%
\expandafter\ifx\csname l@#1\endcsname\relax
\typeout{** WARNING: IEEEtran.bst: No hyphenation pattern has been}%
\typeout{** loaded for the language `#1'. Using the pattern for}%
\typeout{** the default language instead.}%
\else
\language=\csname l@#1\endcsname
\fi
#2}}
\providecommand{\BIBdecl}{\relax}
\BIBdecl

\bibitem{nyt}
\BIBentryALTinterwordspacing
J.~Coscarelli, ``An a.i. hit of fake ‘drake’ and ‘the weeknd’ rattles the music world,'' \emph{The New York Times}. [Online]. Available: \url{https://www.nytimes.com/2023/04/19/arts/music/ai-drake-the-weeknd-fake.html}
\BIBentrySTDinterwordspacing

\bibitem{josan2024ai}
H.~H.~S. Josan, ``Ai and deepfake voice cloning: Innovation, copyright and artists’ rights,'' \emph{Artificial Intelligence}, 2024.

\bibitem{billboard}
\BIBentryALTinterwordspacing
K.~Robinson, ``Ghostwriter, the mastermind behind the viral drake ai song, speaks for the first time,'' \emph{billboard}. [Online]. Available: \url{https://www.billboard.com/music/pop/ghostwriter-heart-on-my-sleeve-drake-ai-grammy-exclusive-interview-1235434099/}
\BIBentrySTDinterwordspacing

\bibitem{fma_dataset}
M.~Defferrard, K.~Benzi, P.~Vandergheynst, and X.~Bresson, ``{FMA}: A dataset for music analysis,'' in \emph{Proc. ISMIR}, 2017.

\bibitem{fma_challenge}
M.~Defferrard, S.~P. Mohanty, S.~F. Carroll, and M.~Salath\'e, ``Learning to recognize musical genre from audio,'' in \emph{Proc. Web Conf.}, 2018.

\bibitem{bogdanov2019mtg}
D.~Bogdanov, M.~Won, P.~Tovstogan, A.~Porter, and X.~Serra, ``The mtg-jamendo dataset for automatic music tagging,'' in \emph{Proc. ICML Machine Learning for Music Discovery Workshop}, 2019.

\bibitem{kim2002singer}
Y.~E. Kim and B.~Whitman, ``Singer identification in popular music recordings using voice coding features,'' in \emph{Proc. ISMIR}, 2002.

\bibitem{fujihara2005singer}
H.~Fujihara, T.~Kitahara, M.~Goto, K.~Komatani, T.~Ogata, and H.~G. Okuno, ``Singer identification based on accompaniment sound reduction and reliable frame selection,'' in \emph{Proc. ISMIR}, 2005.

\bibitem{mesaros2007singer}
A.~Mesaros, T.~Virtanen, and A.~Klapuri, ``Singer identification in polyphonic music using vocal separation and pattern recognition methods,'' in \emph{Proc. ISMIR}, 2007.

\bibitem{fujihara2010modeling}
H.~Fujihara, M.~Goto, T.~Kitahara, and H.~G. Okuno, ``A modeling of singing voice robust to accompaniment sounds and its application to singer identification and vocal-timbre-similarity-based music information retrieval,'' \emph{IEEE Transactions on Audio, Speech, and Language Processing}, 2010.

\bibitem{lagrange2012robust}
M.~Lagrange, A.~Ozerov, and E.~Vincent, ``Robust singer identification in polyphonic music using melody enhancement and uncertainty-based learning,'' in \emph{Proc. ISMIR}, 2012.

\bibitem{hennequin2020spleeter}
R.~Hennequin, A.~Khlif, F.~Voituret, and M.~Moussallam, ``Spleeter: a fast and efficient music source separation tool with pre-trained models,'' \emph{Journal of Open Source Software}, 2020.

\bibitem{defossez2019music}
A.~D{\'e}fossez, N.~Usunier, L.~Bottou, and F.~Bach, ``Music source separation in the waveform domain,'' \emph{arXiv preprint arXiv:1911.13254}, 2019.

\bibitem{sharma2019importance}
B.~Sharma, R.~K. Das, and H.~Li, ``On the importance of audio-source separation for singer identification in polyphonic music,'' in \emph{Proc. Interspeech}, 2019.

\bibitem{hsieh2020addressing}
T.-H. Hsieh, K.-H. Cheng, Z.-C. Fan, Y.-C. Yang, and Y.-H. Yang, ``Addressing the confounds of accompaniments in singer identification,'' in \emph{Proc. IEEE ICASSP}, 2020.

\bibitem{yakura2022self}
H.~Yakura, K.~Watanabe, and M.~Goto, ``Self-supervised contrastive learning for singing voices,'' \emph{IEEE/ACM Transactions on Audio, Speech, and Language Processing}, 2022.

\bibitem{torres2023singer}
B.~Torres, S.~Lattner, and G.~Richard, ``Singer identity representation learning using self-supervised techniques,'' in \emph{Proc. ISMIR}, 2023.

\bibitem{lee2019learning}
K.~Lee and J.~Nam, ``Learning a joint embedding space of monophonic and mixed music signals for singing voice,'' \emph{Proc. ISMIR}, 2019.

\bibitem{kim2021learning}
K.~L. Kim, J.~Lee, S.~Kum, and J.~Nam, ``Learning a cross-domain embedding space of vocal and mixed audio with a structure-preserving triplet loss,'' in \emph{Proc. ISMIR}, 2021.

\bibitem{xie2024fsd}
Y.~Xie, J.~Zhou, X.~Lu, Z.~Jiang, Y.~Yang, H.~Cheng, and L.~Ye, ``Fsd: An initial chinese dataset for fake song detection,'' in \emph{Proc. IEEE ICASSP}, 2024.

\bibitem{zang2024singfake}
Y.~Zang, Y.~Zhang, M.~Heydari, and Z.~Duan, ``Singfake: Singing voice deepfake detection,'' in \emph{Proc. IEEE ICASSP}, 2024.

\bibitem{royo2018disambiguating}
J.~Royo-Letelier, R.~Hennequin, V.-A. Tran, and M.~Moussallam, ``Disambiguating music artists at scale with audio metric learning,'' \emph{Proc. ISMIR}, 2018.

\bibitem{alonso2022music}
P.~Alonso-Jim{\'e}nez, X.~Serra, and D.~Bogdanov, ``Music representation learning based on editorial metadata from discogs,'' in \emph{Proc. ISMIR}, 2022.

\bibitem{kong2024strada}
Y.~Kong, V.-A. Tran, and R.~Hennequin, ``Strada: A singer traits dataset,'' \emph{arXiv preprint arXiv:2406.04140}, 2024.

\bibitem{wilkins2018vocalset}
J.~Wilkins, P.~Seetharaman, A.~Wahl, and B.~Pardo, ``Vocalset: A singing voice dataset.'' in \emph{Proc. ISMIR}, 2018.

\bibitem{zhang2022m4singer}
L.~Zhang, R.~Li, S.~Wang, L.~Deng, J.~Liu, Y.~Ren, J.~He, R.~Huang, J.~Zhu, X.~Chen \emph{et~al.}, ``M4singer: A multi-style, multi-singer and musical score provided mandarin singing corpus,'' \emph{Proc. NeurIPS}, 2022.

\bibitem{demucs2022hybrid}
S.~Rouard, F.~Massa, and A.~D{\'e}fossez, ``Hybrid transformers for music source separation,'' in \emph{Proc. IEEE ICASSP}, 2023.

\bibitem{defossez2021hybrid}
A.~D{\'e}fossez, ``Hybrid spectrogram and waveform source separation,'' in \emph{Proc. ISMIR Workshop on Music Source Separation}, 2021.

\bibitem{ast21b_interspeech}
Y.~Gong, Y.-A. Chung, and J.~Glass, ``Ast: Audio spectrogram transformer,'' in \emph{Proc. Interspeech}, 2021.

\bibitem{chen2020simple}
T.~Chen, S.~Kornblith, M.~Norouzi, and G.~Hinton, ``A simple framework for contrastive learning of visual representations,'' in \emph{Proc. ICML}, 2020.

\bibitem{spijkervet2021contrastive}
J.~Spijkervet and J.~A. Burgoyne, ``Contrastive learning of musical representations,'' \emph{Proc. ISMIR}, 2021.

\bibitem{saeed2021contrastive}
A.~Saeed, D.~Grangier, and N.~Zeghidour, ``Contrastive learning of general-purpose audio representations,'' in \emph{Proc. IEEE ICASSP}, 2021.

\bibitem{ioffe2015batch}
S.~Ioffe and C.~Szegedy, ``Batch normalization: Accelerating deep network training by reducing internal covariate shift,'' in \emph{Proc. ICML}, 2015.

\bibitem{wang2021understanding}
F.~Wang and H.~Liu, ``Understanding the behaviour of contrastive loss,'' in \emph{Proc. IEEE/CVF CVPR}, 2021.

\bibitem{kingma2014adam}
D.~P. Kingma and J.~Ba, ``Adam: A method for stochastic optimization,'' \emph{Proc. ICLR}, 2015.

\bibitem{bishop2006pattern}
C.~M. Bishop, ``Pattern recognition and machine learning,'' \emph{Springer}, 2006.

\bibitem{meseguer2024}
G.~Meseguer-Brocal, D.~Desblancs, and R.~Hennequin, ``An experimental comparison of multi-view self-supervised methods for music tagging,'' in \emph{Proc. IEEE ICASSP}, 2024.

\bibitem{Bertin-Mahieux2011}
T.~Bertin-Mahieux, D.~P. Ellis, B.~Whitman, and P.~Lamere, ``The million song dataset,'' in \emph{Proc. ISMIR}, 2011.

\end{thebibliography}

\end{document}